\documentclass[12pt,epsf]{iopart}
\usepackage{amssymb}
\usepackage{graphicx}
\usepackage{epsfig}

\usepackage{dcolumn}
\usepackage[usenames]{color}
\usepackage{euscript}
\usepackage{mathrsfs}
\usepackage{enumerate}

\newcommand{\be}{\begin{equation}}
\newcommand{\ee}{\end{equation}}
\newcommand{\bel}[1]{\begin{equation}\label{#1}}
\newcommand{\ba}{\begin{eqnarray}}
\newcommand{\ea}{\end{eqnarray}}
\newcommand{\bal}[1]{\begin{eqnarray}\label{#1}}

\begin{document}

\bibliographystyle{apsrev}

\title{Extracting galactic binary signals from the first round of Mock LISA Data Challenges}

\author{Jeff Crowder$^{1,2}$ and Neil J. Cornish$^1$}
\address{$^1$Department of Physics, Montana State University, Bozeman, MT 59717}
\address{$^2$Jet Propulsion Laboratory, California Institute of Technology, Pasadena, CA 91109}

\begin{abstract}

We report on the performance of an end-to-end Bayesian analysis pipeline for detecting and characterizing galactic binary signals in simulated LISA data. Our principal analysis tool is the Blocked-Annealed Metropolis Hasting (BAM) algorithm, which has been optimized to search for tens of thousands of overlapping signals across the LISA band. The BAM algorithm employs Bayesian model selection to determine the number of resolvable sources, and provides posterior distribution functions for all the model parameters. The BAM algorithm performed almost flawlessly on all the Round 1 Mock LISA Data Challenge data sets, including those with many highly overlapping sources. The only misses were later traced to a coding error that affected high frequency sources. In addition to the BAM algorithm we also successfully tested a Genetic Algorithm (GA), but only on data sets with isolated signals as the GA has yet to be optimized to handle large numbers of overlapping signals. 

\end{abstract}

\maketitle

\section{Introduction}

In June of 2006 the first round of the Mock LISA Data Challenges (MLDCs)~\cite{MLDC,MLDC2} was released by the MLDC Taskforce of the LISA International Science Team. The hope of these challenges is to enhance the burgeoning field of gravitational wave data analysis with a focus on the Laser Interferometer Space Antenna (LISA)~\cite{lppa}, to encourage the creation and sharing of ideas, methods, and techniques used by the various groups working on LISA data analysis, and to invite others to join in this interesting field of research. The MLDC also provides a series of blind tests, of increasing complexity, for testing the performance of these various techniques. The first challenge was separated into three areas based on the types of sources whose gravitational waveforms were injected into mock LISA data streams. The first type of source (used in Challenge 1.1) was monochromatic galactic binaries, the second type of source (used in Challenge 1.2) was Schwarzschild massive black hole binaries (MBHBs), and the third type of source (used in Challenge 1.3) was extreme mass ratio inspirals (EMRIs). This work will cover the methods and techniques we used in preparing our entry for Challenge 1.1 involving monochromatic source of gravitational waves.

Challenge 1.1.1 was divided into three parts, each with the signal of a single galactic binary system injected into a mock LISA data stream containing gaussian instrumental noise. This provided a simple test of algorithms across the LISA frequency band. Challenge 1.1.2 contained 20 so-called ``verification binaries'' whose frequency and sky location were known. Challenge 1.1.3 contained the signals of 20 galactic binary systems across the LISA band, with no other information provided to aid the search. Challenge 1.1.4 was the first of the challenges to involve source confusion, where the signals from two or more binaries would overlap and thus possibly affect (confuse) data analysis techniques. Here between $40$ and $60$ (the actual number was unknown) signals were injected into a $15 \mu{\rm Hz}$ band, containing instrument noise. Challenge 1.1.5 was designed to test an algorithm's ability to handle strong source confusion. In this challenge between $30$ and $50$ sources were injected into a $3 \mu{\rm Hz}$ band, containing instrument noise.

Our approach to these challenges was to use two algorithms that we have developed for LISA data analysis to search these various data streams, and extract the seven parameters that describe a monochromatic gravitational wave source: sky location ($\theta, \phi$), frequency ($f$), Amplitude ($A$), polarization ($\psi$), inclination ($\iota$), and initial orbital phase ($\varphi$). The first of these algorithms is the BAM algorithm~\cite{BAM}. A brief overview of the algorithm is given in the next section. The BAM algorithm was used on all levels of Challenge 1.1, though as it was optimized to handle areas where there is significant source overlap,
we will focus our discussion on Challenge 1.1.4 and 1.1.5. The second algorithm we used in these challenges was a Genetic Algorithm (GA)~\cite{genetic}. A brief overview of this algorithm is given in Section \ref{GA_Overview}. As this algorithm has yet to be optimized for regions of high source confusion we used the GA on Challenges 1.1.1, 1.1.2, and 1.1.3. Please note that these two methods of searching were conducted independently of each other. Results garnered from the BAM algorithm were not used to aid the GA searches or vice-versa.

\section{BAM Overview}\label{BAM_Overview}

In this section we give a quick overview of the BAM algorithm, for a more in depth description please see \cite{BAM}.

The BAM algorithm is a variant of the Markov Chain Monte Carlo (MCMC) approach~\cite{metro,haste,gamer} - a powerful set of techniques used to obtain the Posterior Distribution Functions (PDFs) for a model that have been used in analyzing data in many different fields with excellent results. They were first brought to the study of gravitational wave data analysis by Christensen and Meyer~\cite{cm1}. Since then the methods have been adapted to searching for sources in the LISA data stream~\cite{MCMC, MCMC_ed_neil, vecchio_1, MCMC_ed_neil2, vecchio_2, MLDC_Caltech_JPL, MLDC_MT_AEI}. The ability of MCMC techniques to explore large parameter spaces, with multiple sources makes them ideally suited to LISA data analysis. For a more complete review of MCMC methods see~\cite{mcmc_hist, andrieu}.

In an MCMC approach one wants to generate a sample set, $\{ \vec{x} \}$ that corresponds to draws made from the posterior distribution of the system, $p(\vec{\lambda} \vert s)$. To do this one only needs to implement the following simple algorithm.

Beginning at a point, $\vec{x}$, in the parameter space for the system in question, propose a jump to a new point, $\vec{y}$, based on some proposal distribution, $q(\cdot \vert \vec{x})$. The jump is accepted with probability $\alpha = {\rm min}(1,H)$, where $H$ is the Hasting ratio.
\begin{equation}\label{Hastings_ratio}
H = \frac{p(\vec{y}) p(s \vert \vec{y}) q(\vec{x} \vert \vec{y})}
{p(\vec{x}) p(s \vert \vec{x}) q(\vec{y} \vert \vec{x})} \, ,
\end{equation}
Here $p(\vec{x})$ is the prior of the parameters at $\vec{x}$, $q(\vec{x} \vert \vec{y})$ is the value of the proposal distribution for a jump from $\vec{x}$ to $\vec{y}$, and $p(s \vert \vec{x})$ is the likelihood at $\vec{x}$. If the noise is a normal process with zero mean, the likelihood is given by~\cite{sam}: 
\begin{equation}\label{likely}
p(s \vert \vec{x}) = C \exp\left[ -\frac{1}{2} \left( (s - h(\vec{x})) \vert (s - h(\vec{x}))
\right) \right]\, ,
\end{equation}
where the normalization constant $C$ is independent of the signal, $s$, and $(a \vert b )$ denotes the noise weighted inner product
\begin{equation}\label{general_inner_product}
(a \vert b ) = 2 \int_0^\infty \frac{ \tilde{a}^{\ast}(f) \tilde{b}(f) + \tilde{a}(f) \tilde{b}^{\ast}(f)}{S_n(f)} \,df ,
\end{equation}
where $a$ and $b$ are the gravitational waveforms, and $S_n(f)$ is the one-sided noise spectral density. 

Repeated jumps using this algorithm will ensure convergence to the correct posterior for any (non-trivial) proposal distribution~\cite{gilks}. To decrease the time to reach convergence one wants use a proposal distribution (or distributions) that closely models the expected PDF. However, since we do not know the exact form of the posterior a priori, we choose to maintain maximum flexibility in our choice of proposal distributions. We achieve this by using a menu of various proposal distributions, including occasional ``bold'' proposals that attempt large variations in parameter values along with many ``timid'' proposals that attempt small variations in the parameter values of the chain (for a detailed description of some of these proposals see~\cite{MCMC}). This provides a simple means to both search the parameter space and develop the PDFs for the systems.

The BAM algorithm starts with a search phase that is followed by an exploration of the parameter posteriors. During the search phase simulated annealing is used to encourage exploration of the full parameter space. The annealing process effectively smooths the likelihood by the inclusion of a heating term. As the heat is removed the smoothed likelihood surface slowly anneals to the true surface. During this search phase the proposal distributions used to drive the Metropolis-Hastings sampling can be non-Markovian. Thus the search phase of the BAM algorithm is not a true MCMC method. However, when the search phase is complete, and the sampling phase has begun, the proposal distributions that are used are purely Markovian, and the subsequent portion of the chain can be used to properly explore the PDFs.

In our previous work~\cite{BAM} the BAM algorithm used the F-statistic~\cite{fstat} to limit the search to three variables per template, frequency and sky location ($f$, $\theta$, $\phi$). The extrinsic parameters - amplitude, polarization, inclination, and initial orbital phase ($A$, $\psi$, $\iota$, $\varphi_o$) - were then recovered algebraically. While this method is still used as a first approximation for our searches in this work, we have extended the BAM algorithm to a full $7$ parameter search so that we may match the PDFs for all seven parameters. Another update is that we have replaced the low frequency approximation with the rigid adiabatic approximation~\cite{rigida}, using a fast new algorithm developed by Cornish \& Littenberg~\cite{neil_tyson}. The new waveforms give excellent matches to the full LISA Simulator~\cite{lisa_sim} output across the entire LISA band.

\begin{figure}[h]
\includegraphics[angle=270,width=12cm]{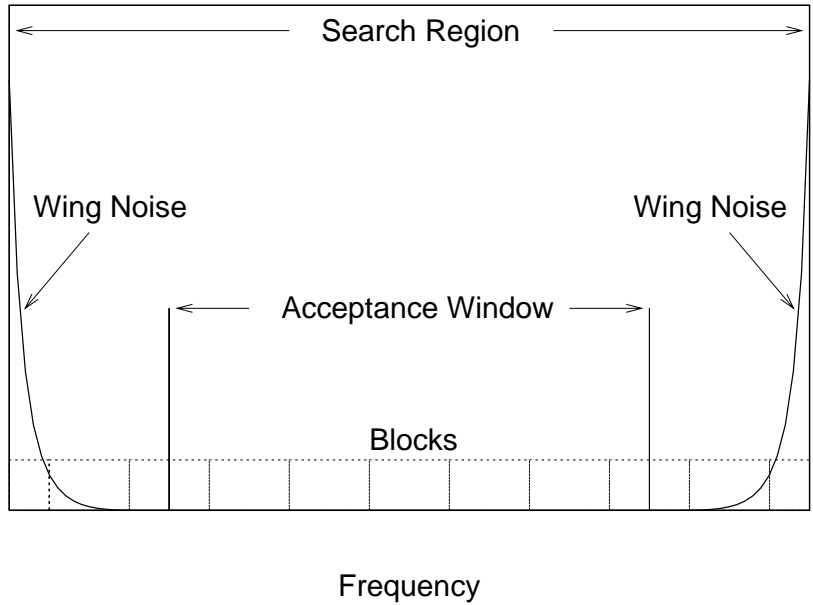}
\caption{\label{Block_MH_figure}A schematic representation of a search region in frequency space, showing
the block structure and some features of the BAM algorithm.}
\end{figure}

The BAM algorithm is optimized to search for multiple, densely packed monochromatic or mildly chirping signals. A key element of the BAM algorithm is blocking. The blocks in the BAM algorithm are small sub-units of the frequency range being searched. As can be seen in Figure~\ref{Block_MH_figure}, which shows a schematic representation of a search region in our BAM algorithm, the search region is broken up into equal sized blocks. The algorithm steps through these blocks sequentially, updating all sources within a given block simultaneously. After all blocks have been updated, they are shifted by one-half the width of a block for the next round of updates. This allows two correlated sources that might happen to be located on opposite sides of a border between two neighboring blocks to be updated together on every other update. It also lessens the number of parameters being updated, which greatly decreases the time needed to calculate non-linear proposal distributions that provide particularly good mixing of the chains.

This blocking allowed for quick and robust searching of isolated data snippets with up to $\sim 100$ templates. The limit on the number of templates is due to the computational cost of the (non-linear) multi-source F-statistic. In order to be able to handle the entire LISA band, we have to break the search up into sub-regions containing a manageable number of sources. This introduced the problem of edge effects from sources whose frequency lay just outside the chosen search region, but deposited power into the search region. To combat the edge effects we introduced ``wings'' and an ``acceptance window.'' The purpose of the wings is to create a buffer between the chosen search region and the regions beyond, so that sources in those outer regions would not adversely affect the ``searchers'' (individual components of the multi-template) in the acceptance window. The search would be run the same as before, but at the end only those searchers inside the acceptance window would be considered as having been found. Searchers that ended up in the wings were discarded (even though many might be perfectly good fits to actual sources). This would allow us step through frequency space, using multiple search regions. But that was not enough. We had to introduce a feature we call ``wing noise'' to attenuate the contribution from the wings in the noise weighted inner product. The wing noise is an increase in the noise spectral density in the wings increasing exponentially from the edge of the acceptance window to the edge of the search region. This keeps searchers from trying to fit to power bleeding into the search region from bright sources nearby. This problem which we call ``slamming'' is also lessened by hierarchically searching the LISA band, finding the brightest sources, and removing the signals of those sources who lay outside the search region, but bleed power into it.

As one can fit the noise in the system, continuing to add more searchers into the search region can steadily increase the value of the likelihood. Thus some other means of deciding what is the best number of searchers must be used. The optimal method would be to calculate evidence ($p_X(s)$) for each model ($X$) given the data ($s$), which is given by:
\begin{equation}\label{marginal}
p_X(s) = \int p(s \vert \vec{\lambda},X) p(\vec{\lambda},X) d\vec{\lambda} \, .
\end{equation}
However, computing this integral is prohibitively expensive for high dimension models. If we assume uniform priors we may use the Laplace approximation estimate:
\begin{equation}
p_X(s) \simeq p(s \vert \vec{\lambda}_{\rm ML},X) \left( \frac{\Delta V_X}{V_X} \right) \, ,
\end{equation}
where $p(s \vert \vec{\lambda}_{\rm ML},X)$ is the maximum likelihood for the model, $V_X$ is the volume of the model's parameter space, and $\Delta V_X$ is the volume of the uncertainty ellipsoid (which can be estimated using a Fisher Information Matrix). It is the $\Delta V_X/V_X$ term that will penalizes models with larger numbers of sources, thus serving as a built in Occam factor. The model with the best (approximate) evidence is selected.

Thus one can divide up the LISA frequency band into multiple search regions, searching each for the resolvable sources that lie within. This allows for searches to be conducted in parallel (after a few pilot searches to identify the brightest of sources that might lead to slamming) on a distributed computing system, reducing the real-world time needed to search the entire LISA band to a matter of days.

\section{GA Overview}\label{GA_Overview}

The GA is an algorithm that uses an evolving solution set whose breeding is based on fitness, as measured by (the logarithm of) the likelihood function, Equation (\ref{likely}). The greater fitness a template (organism) has, the more likely it is to breed. Through breeding and mutation the set of organisms evolves toward an optimal fit to the parameter values of the sources of the gravitational waves in the data stream. In our previous work~\cite{genetic} the GA also used the F-statistic to limit the search space, and again we have extended the search to $7$ parameters for this work. The GA is used in this work to search only for the signals of the isolated binary systems (Challenges 1.1.1 - 1.1.3). This was done because the GA has not yet been fully optimized for high density, multi-source searches. 

The GA that is used in these searches is the Genetic-Genetic Algorithm (GGA) discussed in our previous work. They are considered to be doubly genetic since the mutation rate of the parameters used when creating the next generation is controlled by the algorithm itself, thus optimizing the mutation rate along with the parameter values. In the previous work the Parameter Mutation Rate (PMR) was one value (per organism) that controlled the mutation rate of all the parameters. Here we have implemented distinct PMRs for each of the parameters being searched over (i.e. each organism has three PMRs when an F-statistic based search is run, and seven PMRs when a full parameter search is run). As the GGA is the only type of GA used in these searches, all will be referred to henceforth simply as GAs. The searches conducted here consisted of $10$ organisms per generation with an elitism factor of $0.9$ (i.e. the most fit organism was included in the next generation, and $9$ new organisms were bred to complete the new generation). Also, mild simulated annealing was used to boost exploration of the likelihood surface during the initial phase of the search.

The two breeding patterns used here are known as $1$-point crossover and $n$-point crossover. In the case of $1$-point crossover, breeding is accomplished by splicing the combination of complimentary sections of the binary strings used to represent the parameter values of two parent organisms. For $n$-point crossover each binary digit of the offspring's binary strings is randomly drawn from the corresponding locations in the parent strings. The breeding method was chosen at random for each breeding pair, though as the algorithm progressed the $1$-point crossover was increasingly preferred to the $n$-point crossover as $n$-point crossover breeding enhances exploration of the parameter space, while $1$-point crossover enhances stability after the algorithm has hopefully discovered the peak likelihood values of the space.

\section{Search Results}


\subsection{MLDC 1.1.1}

In these three searches we are seeking the signals from a single binary system injected into a LISA data stream. In the first of the three challenges (1.1.1a) the source was known to have a frequency in the range $[0.9,1.1]~{\rm mHz}$. The source for Challenge 1.1.1b was in the frequency range $[2.9,3.1]~{\rm mHz}$. While the source for Challenge 1.1.1c was in the range $[9,11]~{\rm mHz}$. 

In Challenges 1.1.1a and 1.1.1b the algorithms were allowed to search the entire range for the sources. Exploratory searches of the entire ranges by both the GA and BAM algorithms quickly localized down to the neighborhoods of $1.0627 {\rm mHz}$ and ${3.0004 \rm mHz}$, respectively. Results that were easily verified by viewing the Fourier Transform of the time series data. For Challenge 1.1.1c the Fourier Transform was used to quickly localize the search to the region around $10.5712 {\rm mHz}$.

\begin{figure}
\begin{center}
\epsfig{file=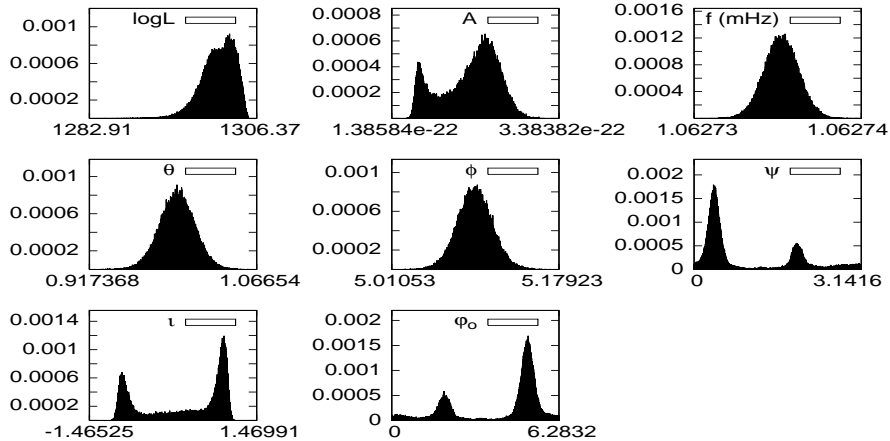,width=6cm,height=12cm,angle=-90}
\end{center}
\caption{\label{MLDC_1.1.1a_histograms}Histograms of the parameter values in the sampling phase of a BAM search of MLDC Challenge Data Set 1.1.1a.}
\end{figure}

The first detailed searches of these localized ranges were performed using the F-statistic versions of the algorithms. The parameter values that were determined by these initial searches were then used as the starting points for the full $7$ parameter searches (with the values obtained by the GA used for further GA searches, and the values obtained by the BAM algorithm used for further BAM searches). The values found by this $7$ parameter searching using GA are quoted in Table~\ref{MLDC_1.1.1_GA_table}, along with the uncertainty estimations for each parameters as determined using a Fisher Information Matrix (FIM) calculated at the recovered parameter values. However, the BAM algorithm included one more run per source to fully map out the PDF for the sources' parameters. These searches used a highly efficient small uniform proposal, able to perform $1,000,000$ steps in less than ten minutes. The parameter values found by the BAM algorithm are shown in Table~\ref{MLDC_1.1.1_BAM_table}, while Figure~\ref{MLDC_1.1.1a_histograms} shows the histograms derived from the PDF exploration of Challenge 1.1.1a (note: this is the mode of marginalized PDF for each the parameter type, not the mode of the full $7N$ dimensional posterior). Also included in Table~\ref{MLDC_1.1.1_BAM_table} are two sets of uncertainty ranges determined a FIM approach and a calculation of the standard deviations of the sources' associated histograms.

One issue to note in the results of this section are highlighted by the multi-modal nature of the histograms in Figure~\ref{MLDC_1.1.1a_histograms}. There is a degeneracy in the polarization and initial phase parameters for monochromatic sources, such that a shift of $\frac{\pi}{2}$ in polarization  with a corresponding shift of $\pi$ in initial phase represents the same physical system. With this taken into account nearly all parameters have been recovered to within $2-\sigma$ as given by the FIM (and the chain standard deviations for the BAM). The notable exceptions being the parameter values for the extrinsic parameters in Challenge 1.1.1c. This difference is most likely due to a model mismatch between the coding of our algorithms and that of the MLDC, possibly caused by implementing the new waveform generation method while the rushing to meet the December deadline. 

\begin{table*}[t]
\caption{\label{MLDC_1.1.1_GA_table}Results of GA searches of the MLDC Challenge Data Sets 1.1.1a, 1.1.1b, and 1.1.1c.}
\begin{tabular}{|l|ccccccc|}
\hline
{ }  & $A$ ($10^{-22}$) & $f$ (mHz) & $\theta$ & $\phi$ & $\psi$ & $\iota$ & $\varphi_0$ \\
\hline 
1.1.1a & { } & { } & { } & { } & { } & { } & { }  \\
\hline 
MLDC Parameters & 2.3900 & 1.062731443 & 0.98064 & 5.0886 & 3.703 & 0.81262 & 4.8951 \\       
GA - Top Organism & 2.5739 & 1.062732487 & 0.99325 & 5.0912 & 3.479 & 0.90351 & 5.2033 \\     
FIM Uncertainties & 0.1783 & 0.000000705 & 0.01451 & 0.0164 & 0.090 & 0.06968 & 0.1956 \\       
\hline                       
1.1.1b & { } & { } & { } & { } & { } & { } & { }  \\
 \hline                      
MLDC Parameters & 0.4805 & 3.000357193 & -0.096411 & 4.6254 & 5.0491 & 1.2404 & 6.1377 \\       
GA - Top Organism & 0.5014 & 3.000356879 & -0.057524 & 4.6219 & 5.0192 & 1.2847 & 6.0208 \\     
FIM Uncertainties & 0.0186 & 0.000000529 & 0.025571 & 0.0051 & 0.0233 & 0.0222 & 0.0705 \\       
\hline                       
1.1.1c & { } & { } & { } & { } & { } & { } & { }  \\
 \hline                      
MLDC Parameters & 0.5841 & 10.57116535 & -0.110638 & 4.6640 & 2.0908 & 0.5182 & 0.964 \\       
GA - Top Organism & 0.2032 & 10.57116504 & -0.124037 & 4.6646 & 4.2757 & 1.5799 & 2.080 \\     
FIM Uncertainties & 0.0185 &  0.00000255 & 0.006139 & 0.0051 & 0.0108 & 0.0446 & 0.270 \\ 
\hline  
\end{tabular}
\end{table*}

\begin{table*}[t]
\caption{\label{MLDC_1.1.1_BAM_table}Results of BAM searches of the MLDC Challenge Data Sets 1.1.1a, 1.1.1b, and 1.1.1c. An asterisk (*) denotes parameter uncertainties from a multimodal histogram.}
\begin{tabular}{|l|ccccccc|}
\hline
  & $A$ ($10^{-22}$) & $f$ (mHz) & $\theta$ & $\phi$ & $\psi$ & $\iota$ & $\varphi_0$ \\
\hline 
1.1.1a & { } & { } & { } & { } & { } & { } & { }  \\
\hline 
MLDC Parameters & 2.390 & 1.06273144 & 0.98064 & 5.0886 & 3.703 & 0.81262 & 4.8951 \\       
Histogram Mode & 2.500 & 1.06273281 & 0.99545 & 5.0969 & 3.507 & 0.89701 & 5.1233 \\ 
Histogram Mean & 2.313 & 1.06273265 & 0.99560 & 5.0943 & 4.041 & 0.21241 & 4.2207 \\ 
FIM Uncertainties & 0.176 & 0.00000070 & 0.01447 & 0.0164 & 0.087 & 0.06805 & 0.19184 \\ 
Chain Std. Dev. & 0.33* & 0.00000070 & 0.01488 & 0.0165 & 0.846* & 0.7402* & 1.5708* \\ 
\hline 
1.1.1b & { } & { } & { } & { } & { } & { } & { } \\
 \hline 
MLDC Parameters & 0.4805 & 3.000357193 & -0.096411 & 4.6254 & 5.0491 & 1.2404 & 6.1377 \\       
Histogram Mode & 0.4923 & 3.000356759 & -0.056613 & 4.6224 & 5.0196 & 1.2828 & 6.0407 \\ 
Histogram Mean & 0.4973 & 3.000356892 & -0.055572 & 4.6223 & 5.0176 & 1.2807 & 6.0236 \\ 
FIM Uncertainties & 0.0187 & 0.000000529 & 0.025737 & 0.0051 & 0.0237 & 0.0226 & 0.0710 \\ 
Chain Std. Dev. & 0.0188 & 0.000000546 & 0.026948 & 0.0052 & 0.0235 & 0.0229 & 0.0727 \\ 
\hline 
1.1.1c & { } & { } & { } & { } & { } & { } & { } \\
 \hline 
MLDC Parameters & 0.5841 & 10.57116535 & -0.11064 & 4.6640 & 2.0908 & 0.51822 & 0.9642 \\       
Histogram Mode & 0.4038 & 10.57116568 & -0.11906 & 4.6654 & 1.1294 & 0.48558 & 1.9583 \\ 
Histogram Mean & 0.4227 & 10.57116579 & -0.11758 & 4.6651 & 1.1050 & 0.20026 & 2.2932 \\ 
FIM Uncertainties & 0.1386 & 0.000000561 & 0.01295 & 0.0023 & 0.1507 & 0.70556 & 0.3018 \\ 
Chain Std. Dev. & 0.0249 & 0.000000343 & 0.00570 & 0.0006 & 0.1342 & 0.28290 & 0.2635 \\ 
 \hline 
\end{tabular}
\end{table*}

As Challenges 1.1.2 and 1.1.3 were similar to Challenge 1.1.1, in that they featured essentially isolated sources, we will not give a review of those results here, but instead direct the interested reader to the evaluation of all of the entries to Round 1 of the MLDC in this volume \cite{Arnaud-summary}.

\subsection{MLDC 1.1.4}

This challenge involved a search over a restricted frequency range ($\Delta f = 15 \mu{\rm Hz}$) containing a large number of injected signals ($n = 45$). As the GA has not been optimized for searching such extended frequency ranges with high numbers of sources, only the BAM algorithm was used in searching this data stream. 

Here the BAM algorithm was used hierarchically, initially searching for a much smaller number of sources than was expected. The purpose  of the initial run with few searchers is to pick off the brightest sources in the data stream so that in the next step, when the data is divided up in to separate frequency windows, the bright sources from one window can be accounted for in nearby windows. Thus the initial run was performed searching the entire frequency range of the data looking for only $5$ sources. For the succeeding runs, the data stream was separated into five snippets with windows of acceptance $3 \mu{\rm Hz}$ in width and wings (acting as a buffer against edge effects) $0.91 \mu{\rm Hz}$ in width (please see \cite{BAM} for a full description on exactly how the wings provide this buffer and other aspects of multiple source searching using BAM). Sources that are discovered in the windows of acceptance are kept, those found in the wings are disregarded. In the hierarchical searches, first one searcher is sent off to look for a source, and after a set number of steps in the chain the value of the evidence is recorded and a second searcher is added to the search. If, after a specified number of steps the evidence does not show that the fit has improved enough the search is ended and the results of the second searcher are discarded. If the evidence warrants keeping the second searcher, then a third searcher is added and the search continues (up to a maximum of $5$ searchers in this particular run).

\begin{figure}[h]
\begin{center}
\epsfig{file=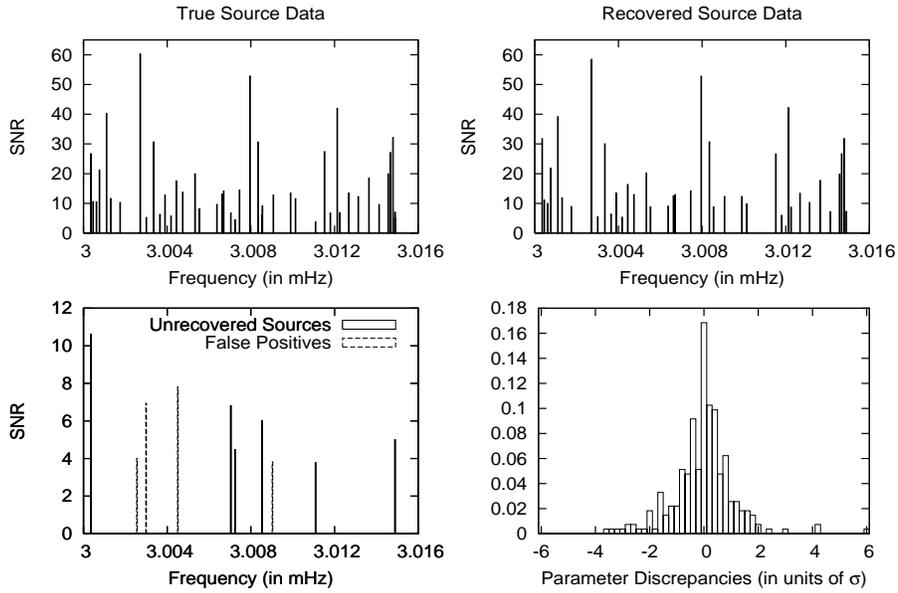,width=8cm,height=12cm,angle=-90}
\end{center}
\caption{\label{MLDC_1.1.4_figure}Plots detailing the results of a search of the LISA data stream for Challenge data set 1.1.4 of the Mock LISA Data Challenge. The upper two plots show the frequencies and SNRs for the true and recovered sources (the width of the bars is $1$ frequency bin). The lower left plot gives the frequencies and SNRs of the sources that were not recovered as well as the false positives. The lower right plot shows the discrepancies between the true and recovered parameters in units of the parameter variances.}
\end{figure}

In the next step the searches were performed on the divided data stream, which was separated into $5$ search regions. There was a search for up to $6$ sources in each of the $5$ windows. After this step, the algorithm had isolated $26$ candidate sources. The next run over the divided data stream (using the same $5$ windows), was a search for up to $8$ sources in each of the $5$ windows. To improve the chances of finding sources, $5$ different starting seeds for each window were run in parallel on a supercomputer cluster. These runs consisted of up to $300,000$ steps in the chains (all runs were stopped by the evidence criteria before reaching this mark). Results from parallel chains were merged and duplicates discarded. After this step, the algorithm had isolated $43$ candidate sources, $39$ of which were considered to be recovered (i.e. they had at most $1$ of the $7$ parameters greater than $2-\sigma$ discrepant from the listed MLDC parameters). The four false positives represent searchers whose parameters do not match this criteria. Figure~\ref{MLDC_1.1.4_figure} gives a sampling of the results of this search. One thing to note is of the sources that were not recovered, the brightest (SNR $= 10.6$) shared an identical frequency with a brighter (SNR $= 26.7$) recovered source, and which was located only $\sim 5^\circ$ on the sky from the unrecovered source. This extreme similarity in sources represents one of the hardest challenges for LISA data analysis. 

\subsection{MLDC 1.1.5}

This challenge involved a search over a smaller frequency range than Challenge 1.1.4, ($\Delta f = 3 \mu{\rm Hz}$), yet it still contained a fairly large number of sources ($n = 33$). It was designed to test an algorithm's ability to deal with very high source densities (more than $1$ source every $3$ frequency bins). As in Challenge 1.1.4, only the BAM algorithm was used in searching this data stream. 

The first hierarchical search looked for $8$ sources over the entire data stream. In the next search, the data stream was divided into six windows which were each searched for up to $8$ sources. These runs were performed using $3$ different starting seeds. The evidence was used to stop the search process if increasing the number of searchers did not provide enough benefit to the overall fit before that limit. While most stopped after reaching $8$ sources, many of these were found in the wings so that the number of candidates found in the acceptance windows of the searches only numbered $27$. So at the end of this round there were $27$ candidate sources. Another run was attempted in which the algorithm could search for up to $12$ sources per window. This run did not produce any more viable candidates, in part due to the analysis method used to study the chains. The candidate source parameters were chosen by taking the average values of the chains in that sampling phase of the search. However, if there are other sources nearby the chain picks up secondary modes, and the average ends up being a blend of the two candidate sources. This is more of a problem for low SNR sources, and presents a limit of the effectiveness of the BAM algorithm when using chain averaging. We intend to move to choosing the candidate sources by either analyzing the modes of the histograms of the output chains or by using the maximum a posteriori (MAP) value to determine parameter values for candidate sources in the data stream. 

\begin{figure}[h]
\begin{center}
\epsfig{file=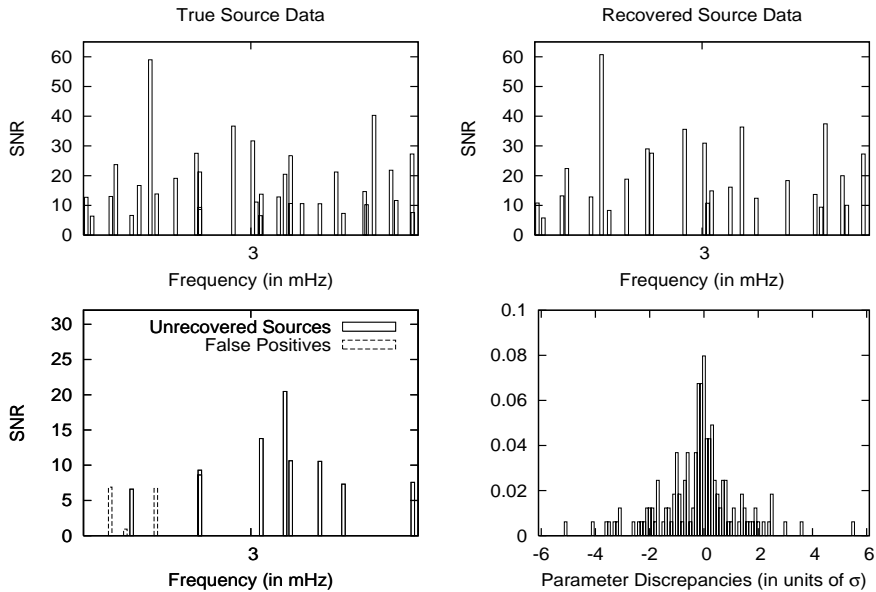,width=8cm,height=12cm,angle=-90}
\end{center}
\caption{\label{MLDC_1.1.5_figure}Plots detailing the results of a search of the LISA data stream for Challenge data set 1.1.5 of the Mock LISA Data Challenge. The upper two plots show the frequencies and SNRs for the true and recovered sources (the width of the bars is $1$ frequency bin). The lower left plot gives the frequencies and SNRs of the sources that were not recovered as well as the false positives. The lower right plot shows the discrepancies between the true and recovered parameters in units of the parameter variances.}
\end{figure}

Figure~\ref{MLDC_1.1.5_figure} gives a sampling of the results of this search. As with Challenge 1.1.4, this challenge also contained sources with identical frequencies. In this case there was a pair of binaries sharing one identical frequency (both located within $5^\circ$ of each other on the sky) as well as a triplet of binaries sharing another identical frequency (all located within $6^\circ$ of each other on the sky). The BAM algorithm was only able to recover $2$ of these $5$ sources before the deadline, and using averaging of the chains. Two more of the unrecovered sources had a second source located less than one-quarter of one frequency bin from a recovered source. We hope to be able to separate such sources with the next version of the algorithm.

\section{Conclusion}

The GA and BAM algorithms performed well in these first MLDCs. While the GA has yet to be optimized for multi-source searching, it was able to recover the parameters of the isolated sources quite well. The BAM algorithm was especially suited to the multi-source cases in Challenges 1.1.4 and 1.1.5. It recovered more than $85 \%$ of the sources present in Challenge 1.1.4, and more than $70 \%$ of the sources present in the extremely source-dense data stream of Challenge 1.1.5. Also, with its ability to partition of the frequency band, the BAM algorithm appears to be ready to tackle the second round of MLDCs~\cite{Arnaud-second}, which will include the gravitational wave signals from $25,000,000^+$ galactic binary systems. However, there is still room for improvement. Some suggestions have been given, particularly moving away from using chain averaging for parameter determination. Another improvement to the BAM that will be paramount when attempting the next round of MLDCs is introduction of the noise level as a searchable parameter. This should allow the algorithm to remove the signals to near the confusion limit.

\ack
This work was supported at MSU by NASA Grant NNG05GI69G. The searches described herein were carried out using the computing facilities of the Jet Propulsion Laboratory, California  Institute of Technology, under a contract with the National Aeronautics and Space Administration.

\end{document}